\documentclass[11pt]{article}
\usepackage{moriond,epsfig}

\def\be{\begin{equation}}
\def\ee{\end{equation}}
\def\bea{\begin{eqnarray}}
\def\eea{\end{eqnarray}}

\begin{document}
\vspace*{4cm}
\title{Probing dark matter distribution with gravitational lensing and stellar dynamics}

\author{Tommaso Treu \& L\'eon V.~E.~Koopmans}

\address{California Institute of Technology \\ Astronomy 105-24 \\ Pasadena, CA 91101, USA}

\maketitle\abstracts{The Lenses Structure and Dynamics (LSD) Survey
aims at measuring the luminous and dark matter distribution of high
redshift early-type galaxies by combining gravitational lens analysis
with newly determined velocity dispersion profiles. Here we describe
the first results from the LSD survey, a measurement of the internal
structure of the lens galaxy in MG2016+112 ($z=1.004$). The relevance
of this measurement to the cosmological model is briefly discussed, in
particular in the context of E/S0 galaxy formation and evolution, the
existence and shape of universal dark matter profiles, and the
measurement of the Hubble constant from gravitational time delay.}

\section{Introduction}

Little is known about the distribution of mass in early-type galaxies
(E/S0). In fact, the paucity of kinematic tracers at large radii
(beyond the effective radius R$_e$, i.e. beyond the region dominated
by stellar mass) severely limits the information that can be inferred
by dynamical studies, In the local Universe, particularly accurate and
extensive data have provided evidence for the existence of dark-matter
halos in E/S0, with constraints on the shape of the dark matter
distribution. At higher redshifts, galaxy-galaxy lensing and strong
lensing have provided independent confirmation for the existence of
dark halos, and some information on the mass distribution.

The aim of the Lenses Structure and Dynamical (LSD) Survey is to
measure the luminous and dark matter distribution of E/S0, by
combining lensing and kinematic constraints. The combination of these
two diagnostics provides independent constraints on the same physical
length-scales, effectively removing degeneracies inherent to each
method alone (such as that between anisotropy and mass in kinematic
studies). In particular, this technique allows us to study in detail
the distribution of luminous and dark matter in E/S0 at significant
look-back time (up to $\sim 8$ Gyrs) with accuracy comparable to
studies in the local Universe. In practice, we are measuring (using
the Echellette Spectrograph and Imager ESI at the 10m Keck-II
Telescope) velocity dispersion and streaming motion profiles of a
sample of 11 lens E/S0 ($0<z<1$), chosen to be relatively clean
systems, covering uniformly the redshift range, and spanning the
widest possible range in masses (a more detailed description of the
survey is given in Treu \& Koopmans \cite{TK02}, hereafter TK02).

The relevance of this study to this conference is two-fold. First of
all, as described in the opening review by Jim Peebles, the formation
of early-type galaxies is a very controversial but interesting issue
because it provides a crucial observational constraint on cosmological
models. Most of the studies on E/S0 at intermediate and high redshift
so far were concerned with the evolution of the stellar
populations. The LSD Survey provides additional valuable and detailed
information on the internal structure of high redshift E/S0 and hence
on its evolution. Second, Cold Dark Matter (CDM) cosmological
simulations find that dark matter halos have a universal density
profile (see the next section), characterized by a cuspy inner
slope. This prediction has been challenged by the observation of soft
cores in local dwarf and low surface brightness galaxies (but see
review talk by Simon White). The LSD Survey provides a completely
independent determination of the shape of dark matter halos in massive
E/S0 galaxies-- at different mass scales, physical conditions, and
look-back time.

\section{A Case study: MG2016+112 at $z=1$}

In this contribution we focus in particular on the first object
analyzed by our survey. The lens galaxy in MG2016+112 is the most
distant spectroscopically confirmed lens ($z=1.004$) and therefore
offers the possibility of exploring the internal structure of an E/S0
at a significant look back-time ($\sim$ 8 Gyr in the assumed cosmology
$\Omega_m=0.3$, $\Omega_{\Lambda}=0.7$, H$_0=65$
kms$^{-1}$Mpc$^{-1}$). In addition, the Einstein radius is almost five
times the effective radius (see below) and therefore the mass profile
is constrained out to the largest radii probed even in local E/S0,
yielding a robust determination of the slope of the total mass
distribution. Due to the extreme faintess and distance of MG2016+112,
only a luminosity-weighted velocity dispersion was measured, and
therefore no tight constraints on the orbital structure of the system
could be placed (see Koopmans \& Treu\cite{KT02b} for the analysis of
a system with spatially resolved kinematics and constraints on the
velocity ellipsoid).

\subsection{Observations and modeling}

The lens galaxy in MG2016+112 was imaged with the Hubble Space
Telescope. Surface photometry (Koopmans \& Treu\cite{KT02a}; hereafter
TK02a) shows that the surface brightness profile is well-fit by a de
Vaucouleurs profile with effective radius R$_e=0.31''\pm0.06''$,
i.e. $2.7\pm0.5$ kpc.
\begin{figure}
\psfig{figure=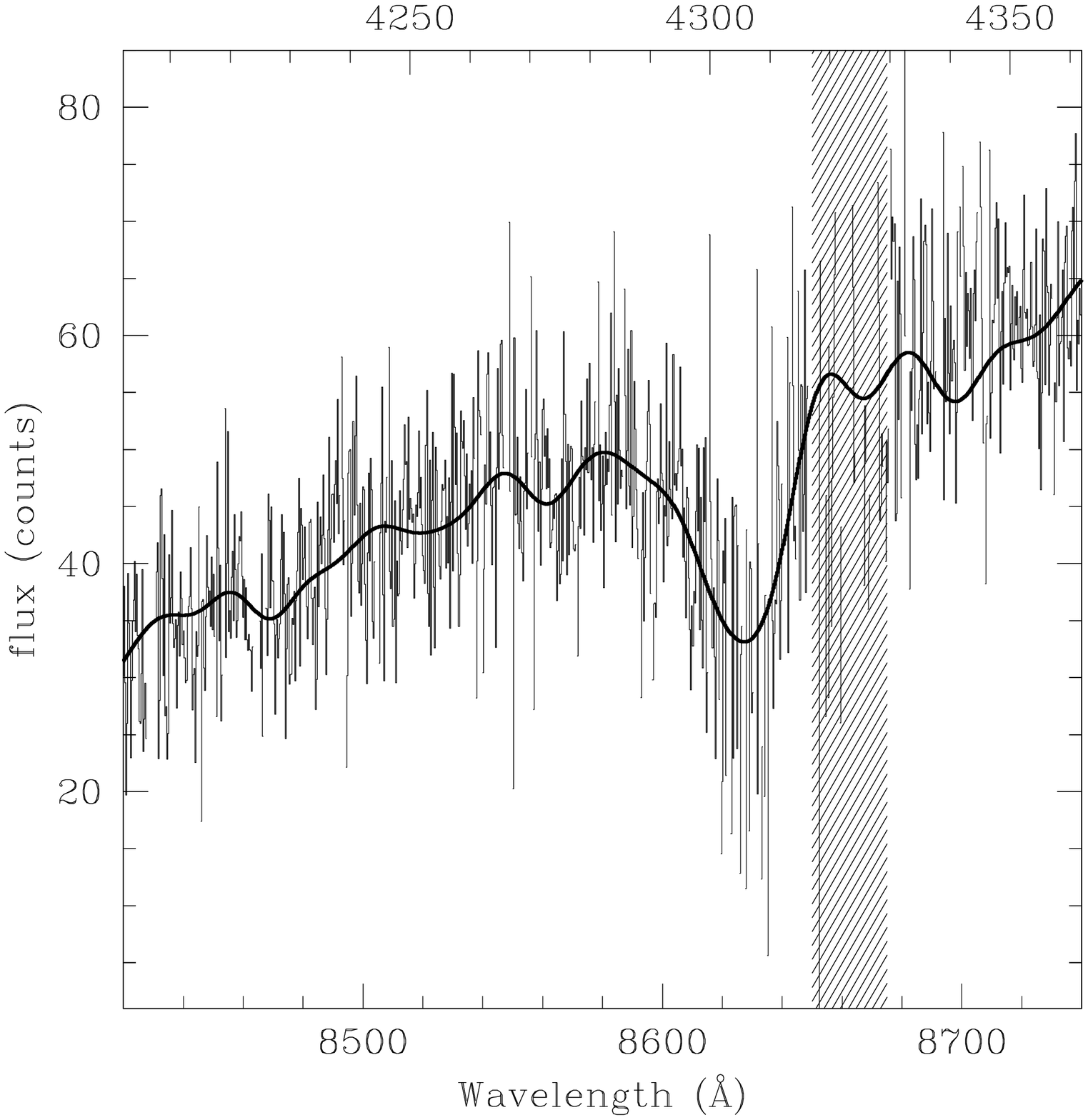,height=2.5in}
\psfig{figure=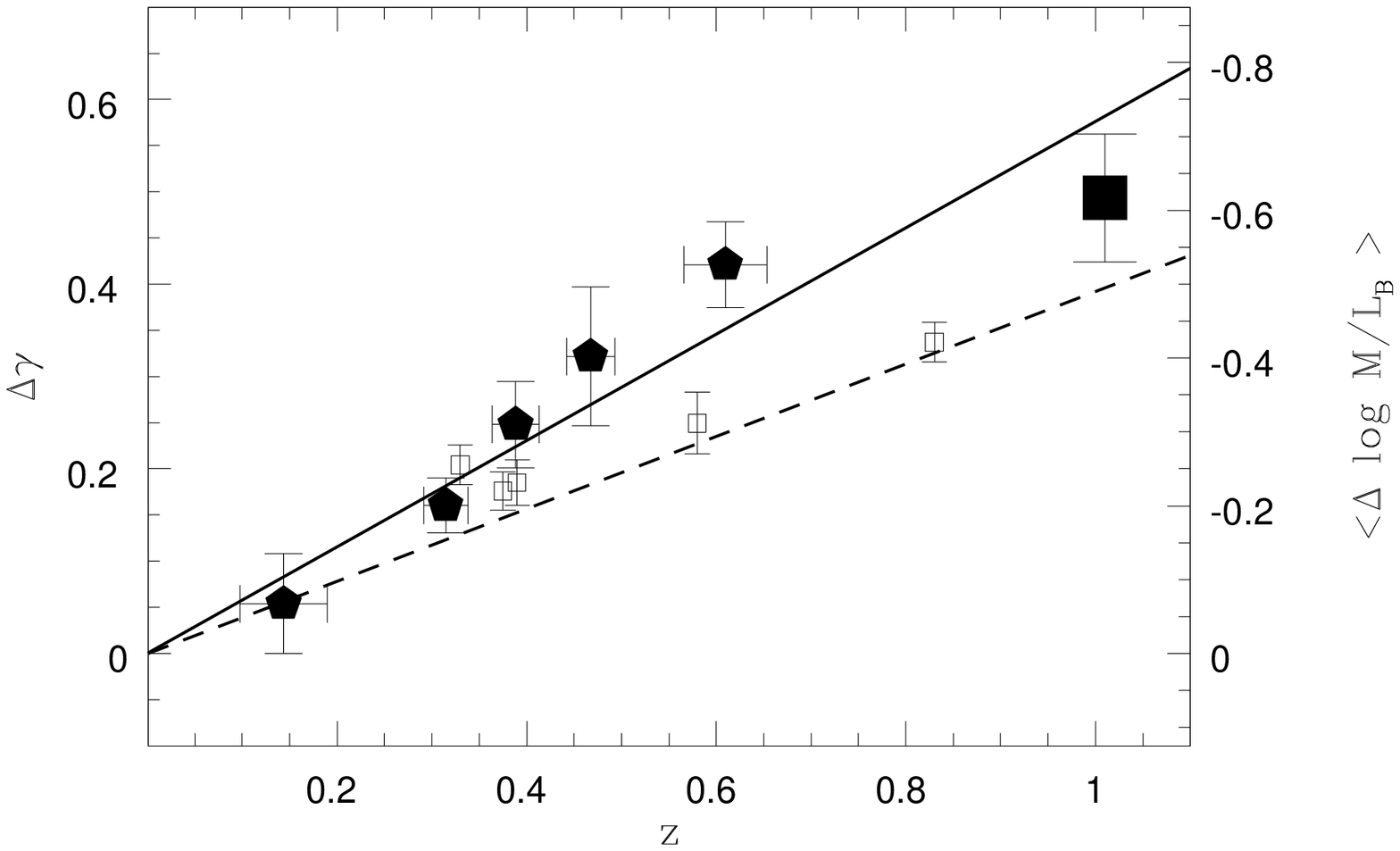,height=2.5in}
\caption{Left: the G-band absorption feature in the spectrum of
MG2016+112. The rest-frame wavelength scale is shown on the top axis
for reference. The smooth solid line is the best-fitting template
(spectral-type G4--III) broadened to the measured velocity dispersion
$\sigma_{\rm ap}=304\pm27$ km\,s$^{-1}$. The dashed region is affected
by sky-subtraction residuals and was omitted in the fit. Right:
Evolution of the M/L as inferred from the evolution of the FP. Filled
pentagons and open squares represent field and cluster measurements,
respectively, while the thick full and dashed lines indicate linear
fits to the M/L evolution for field and cluster E/S0 (see Treu et
al. 2002 for details). The large filled square at $z=1.004$ indicates
the M/L evolution of MG2016+112. \hspace{3cm}
\label{fig:MG2016a}}
\end{figure}
\noindent
We observed MG2016+112 using ESI on the W.M.~Keck--II Telescope during
four consecutive nights (23--26 July, 2001), with a total integration
time of 8.5\,hrs. The region around the absorption G-band (Fig~1) was
used to derive a velocity dispersion of $\sigma_{\rm
ap}=304\pm27$\,km\,s$^{-1}$ inside an effective circular aperture with
radius $0.65''$, corresponding to a central velocity dispersion
(i.e. measured in a $R_e/8$ aperture) of $\sigma =
328\pm32$~km\,s$^{-1}$. Details of the measurement are given in KT02a.

We can use these measurements to determine the offset of the galaxy
with respect to the local Fundamental Plane (Djorgovski \& Davis
\cite{DD87} Dressler et al. \cite{D87}) and therefore measure the
evolution of the effective mass to light ratio (Figure 1 right panel;
see Treu et al.\cite{T01} and KT02a for discussion). Assuming, as in
TK02, that the evolution of the effective mass to light ratio is equal
to the evolution of the stellar mass to light ratio, we can use this
measurement to predict the stellar mass-to-light ratio of MG2016+112
finding $M_*/L_B=1.8\pm0.7$ in solar units.

Finally, the lens configuration provides an accurate measurement of
the mass enclosed by the Einstein Radius (13.7 kpc),
$M_E=1.1\times10^{12} M_{\odot}$ (Koopmans et al.\ \cite{K02}).

The galaxy mass distribution is modeled as a superposition of two
spherical components, one for the luminous stellar matter and one for
the dark-matter halo. The luminous mass distribution is described by a
Hernquist (1990) model, which provides a good analytical approximation
to the luminous component of early-type galaxies. The dark-matter
distribution is modeled as
\begin{equation}
\rho_d(r)=\frac{\rho_{d,0}}{(r/r_b)^{\gamma}(1+(r/r_b)^2)^{(3-\gamma)/2}}
\label{eq:DM}
\end{equation}
which closely describes a Navarro, Frenk \& White\cite{NFW} profile for
$\gamma=1$, and has the typical asymptotic behavior at large radii
found from numerical simulations of dark matter halos $\propto r^{-3}$
(e.g. Ghigna et al.\ 2000). See TK02 for further discussion of this
model.

In addition, we assume an Osipkov-Merritt parametrization of the
anisotropy $\beta$ of the luminous mass distribution
$\beta(r)=1-\frac{\sigma^2_{\theta}}{\sigma_{r}^2}=\frac{r^2}{r^2+r^2_i}$,
where $\sigma_{\theta}$ and $\sigma_{r}$ are the tangential and radial
component of the velocity dispersion and $r_i$ is called the
anisotropy radius (Osipkov\cite{O79}; Merritt \cite{M85a,M85b}). The
line-of-sight velocity dispersion is found solving the spherical Jeans
equation. Finally, the luminosity-weighted average within the
spectroscopic aperture is computed for comparison with the
observations.

\subsection{Results}

The effective radius and the total mass ($M_E$) within the Einstein
radius are fixed by the observations. This leaves four free parameters
in the model: the inner slope ($\gamma$) of the dark matter halo, the
length scale of the dark matter component ($r_b$), the mass-to-light
ratio of the luminous component (M$_*/L_B$) and the anisotropy radius
($r_i$). The effect of changing $r_i$ on the model velocity dispersion
is marginal and therefore $r_i$ can be fixed to $R_e$. Also, according
to the CDM simulations given in Bullock et al.\ \cite{B01}, the scale
radius $r_b$ is much larger than the Einstein Radius, so we will
consider this limit. Hence effectively there are two free parameters
which we can constrain with our observations. The results are
discussed in detail in KT02 and TK02, here we only briefly summarize
the highlights: (i) MG2016+112 is a massive E/S0 and a dark cluster as
suggested by Hattori et al.~\cite{H97} is therefore not necessary to
reproduce the image separation; (ii) the offset from the local FP
shows that the evolution of M/L for MG2016+112 is intermediate between
the extrapolation of the evolution found for cluster and field samples
(Fig.1), and with the independent estimate from our two-component
(dark+luminous) dynamical model; (iii) dark matter contributes more
than 60\% (99\%~CL) of the total mass within the Einstein Radius; (iv)
the effective slope of the {\sl total} mass distribution is very close
to isothermal, i.~e. $\rho\propto r^{-\gamma'}$ with
$\gamma'=2.0\pm0.1\pm0.1$, which could be interpreted as a possible
indication of (incomplete) violent relaxation; (v) the inner slope
$\gamma$ of the dark matter halo is flatter than isothermal
($\gamma<2.0$; 95 \% CL); (vi) using a simple adiabatic contraction
model (Blumenthal et al.\ \cite{B86}) we compute the inner slope,
$\gamma_i$, before baryonic collapse and find that $\gamma_i<1.4$
(68\% CL), thus marginally inconsistent with high resolution CDM
numerical simulations (Moore et al.\cite{M98}).

Note that the uncertainty on the slope of total mass profile is the
main source of uncertainty in the determination of H$_0$ from
gravitational time-delays (e.g Koopmans \& Fassnacht
\cite{KF99}). Therefore, if what we find for MG2016 -- i.e. that the
total mass profile is almost perfectly isothermal -- is found also for
other E/S0 (see e.g. KT02b), this would lead to significant
improvement in the accuracy of H$_0$ determination from gravitational
time-delays.

\section*{Acknowledgments}

We acknowledge financial support from NSF and HST grants
(AST--9900866; STScI--GO 06543.03--95A; STScI-AR-09222).

\end{document}